\documentclass{article}

\def\bSig\mathbf{\Sigma}

\usepackage{color}
\usepackage{bm}

\usepackage[figuresright]{rotating}

\usepackage{authblk}

\usepackage{lscape}
\usepackage{graphicx}
\usepackage[figuresright]{rotating}
\usepackage{bm}

\usepackage{amsmath}

\usepackage{tikz}
\usetikzlibrary{graphs}
\usetikzlibrary{arrows,backgrounds,positioning,fit,calc, intersections, through}
\usetikzlibrary{decorations.pathmorphing,petri}

\title{A matrix-based method of moments for fitting multivariate network meta-analysis models \\with multiple outcomes \\and  random inconsistency effects}
\author[1,*]{Dan Jackson}
\author[2]{Sylwia Bujkiewicz}
\author[1]{Martin Law}
\author[3]{\\Richard D Riley}
\author[1,4]{Ian White}

\affil[1]{MRC Biostatistics Unit, Cambridge, UK}
\affil[*]{email: daniel.jackson@mrc-bsu.cam.ac.uk}
\affil[2]{Biostatistics Research Group, Department of Health Sciences, University of Leicester, UK}
\affil[3]{Centre for Prognosis Research, Research Institute for Primary Care and Health Sciences, University of Keele, UK}
\affil[4]{MRC Clinical Trials Unit at University College London, UK}

\begin{document}

\begin{titlepage}
\maketitle

\begin{abstract}
Random-effects meta-analyses are very commonly used in medical statistics. Recent methodological developments include  multivariate (multiple outcomes) and network (multiple treatments)  meta-analysis. Here we provide a new model and corresponding estimation procedure for multivariate network meta-analysis, so that multiple outcomes and treatments can be included in a single analysis. Our new multivariate model is a direct extension of a univariate model for network meta-analysis that has recently been proposed.   We allow two types of unknown variance parameters in our model, which represent between-study heterogeneity and inconsistency. Inconsistency arises when different forms of direct and indirect evidence are not in agreement, even having taken between-study heterogeneity into account. However the consistency assumption is often assumed in practice and so we also explain how to fit a reduced model which makes this assumption. Our estimation method extends several other commonly used methods for meta-analysis, including the method proposed by DerSimonian and Laird (1986).  We investigate the use of our proposed methods in the context of  a real example.
\end{abstract}
\end{titlepage}


\label{firstpage}


%
%

%
%
\newpage

 \section{Introduction}
Meta-analysis, the statistical process of pooling the results from separate studies, is commonly used in medical statistics and now requires little introduction. The univariate random-effects model  is often used for this purpose. This model has recently been extended to the multivariate (multiple outcomes; Jackson {\it et al.}, 2011) and network (multiple treatments; Lu and Ades, 2004) meta-analysis settings. In a network meta-analysis,  more than two treatments are included in the same analysis. The main advantage of network meta-analysis is that, by using indirect information contained in the network, more precise and coherent inference is possible, especially when direct evidence for particular treatment comparisons is limited. Here we describe a new model that extends the random-effects modelling framework to the multivariate network meta-analysis setting, so that both multiple outcomes {\it and} multiple treatments may be included in the same analysis.

Other multivariate extensions of univariate methods for network meta-analysis have previously been proposed. For example,  Achana {\it et al}. (2014)  analyse multiple correlated outcomes in multi-arm studies in public health. Efthimiou  {\it et al}. (2014) propose a model for the joint modelling of odds ratios on multiple endpoints. Efthimiou {\it et al}. (2015) develop another model that is a network extension of an alternative multivariate meta-analytic model that was originally proposed by Riley {\it et al}. (2008). A network meta-analysis of multiple outcomes with individual patient data has also been proposed by Hong {\it et al}. (2015) under both contrast-based and arm-based parameterizations, and Hong {\it et al}. (2016) develop a Bayesian framework for multivariate network meta-analysis. These  multivariate network meta-analysis models are based on the assumption of consistency in the network, extending the approach introduced by Lu and Ades (2004). In contrast to these previously developed  methods, the method proposed here relaxes the consistency assumption.  This assumption is sometimes found to be false across the entire network (Veroniki {\it et al.}, 2013). We model the inconsistency using a design-by-treatment interaction, so that different forms of direct and indirect evidence may not agree, even after taking between-study heterogeneity into account. However we assume that the design-by-interaction terms follow normal distributions, and so conceptualise inconsistency as another source of random variation. This allows us to achieve the dual aim of estimating meaningful treatment effects whilst also allowing for inconsistency in the network.

 Although we allow inconsistency in the network, we propose a relatively simple model. Our preference for a simple model is because the between-study covariance structure is typically hard to identify accurately in multivariate meta-analyses  (Jackson {\it et al.}, 2011) and also because network meta-analysis datasets are usually small (Nikolakopoulou {\it et al.}, 2014). The new model that we propose for multivariate network meta-analysis is a direct generalisation of the univariate network meta-analysis model proposed by   Jackson {\it et al.} (2016), which is a particular form of the design-by-treatment interaction model (Higgins {\it et al.}, 2012). In addition to proposing a new model for multivariate network meta-analysis, we also develop a corresponding new estimation method. This estimation method is based on the method of moments and extends a wide variety of related methods. In particular, we  extend the estimation method described by DerSimonian and Laird (1986) by directly extending the matrix based extension of DerSimonian and Laird's estimation method for multivariate meta-analysis (Chen {\it et al.}, 2012; Jackson {\it et al.}, 2013).  We adopt the usual two-stage approach to meta-analysis, where the estimated study-specific treatment effects (including the within-study covariance matrices) are computed in the first stage. We give some information about how this first stage is performed but our focus is the second stage, where the meta-analysis model is fitted.

 The paper is set out as follows. In section 2, we briefly describe the univariate model for network meta-analysis to motivate our new multivariate network meta-analysis model in section 3. We present our new estimation method in section 4 and we apply our methods to a real dataset in section 5. We conclude with a short discussion in section 6.

\section{A univariate network meta-analysis model}
Here we describe our univariate modelling framework for network meta-analysis (Jackson {\it et al.}, 2016; Law {\it et al.}, 2016).  Without loss of generality, we take treatment A as the reference treatment for the network meta-analysis. The other treatments are B, C, {\it etc.} We take the design $d$ as referring only to the set of treatments compared in a study.  For example, if the first design compares treatments A and B only, then $d=1$ refers to two-arm studies that compare these two treatments. We define $t$ to be the total number of treatments included in the network, and $t_d$ to be the number of treatments included in design $d$. We define $D$ to be the number of different designs,  $N_d$ to be the number of studies of design $d$, and $N = \sum\limits_{d=1}^{D} N_d$ to be the total number of studies. We will use the word  `contrast' to refer to a particular treatment comparison or effect in a particular study, for example the `AB contrast' in the first study.

We model the estimated relative treatment effects, rather than the average outcomes in each arm, and so perform contrast based analyses. We define ${\bm Y}_{di}$ to be the $c_d \times 1$ column vector of estimated relative treatment effects from the $i$th study of design $d$, where $c_d=t_d-1$. We define $n_d = N_d c_d $ to be the total number of estimated treatment effects that design $d$ contributes to the analysis, and  $n = \sum\limits_{d=1}^{D} n_d$ to be the total number of estimated treatment effects that contribute to the analysis. To specify the outcome data ${\bm Y}_{di}$, we  choose a baseline treatment in each design $d$. The entries of ${\bm Y}_{di}$ are then the estimated  effects of the other $c_d$ treatments  included in design $d$ relative to this baseline treatment. For example, if we  take $d=2$ to indicate the `CDE design' then $c_2=2$. Taking C as the baseline treatment for  this design, the two entries of the  ${\bm Y}_{2i}$ vectors are the estimated relative effects of treatment D compared to C and of treatment E compared to C. For example, the entries of the ${\bm Y}_{di}$ could be estimated log-odds ratios or mean differences.

We use normal approximations  for the within-study distributions.  We define ${\bm S}_{di}$ to be the $c_d \times c_d$ within-study covariance matrix corresponding to  ${\bm Y}_{di}$. We treat all ${\bm S}_{di}$ as fixed and known in analysis.  Ignoring the uncertainty in the ${\bm S}_{di}$ is acceptable provided that the studies are reasonably large and is conventional in meta-analysis, but this approximation is motivated by pragmatic considerations because this greatly simplifies the modelling. We do not impose any constraints on the form of ${\bm S}_{di}$ other than they must be valid covariance matrices. The lead diagonal entries of the ${\bm S}_{di}$ are within-study variances that can be calculated using standard methods. Assuming that the studies are composed of independent samples for each treatment, the other entries of the ${\bm S}_{di}$ are calculated as the variance of the average outcome (for example the log odds or the sample mean) of the baseline treatment.

We define  $\delta_1^{AB}$, $\delta_1^{AC}$, $\cdots$, $\delta_1^{AZ}$, where Z is the final treatment in the network, to be treatment effects relative to the reference treatment A, and call them basic parameters (Lu and Ades, 2006). We use the subscript 1 when defining the basic parameters to emphasise that they are treatment effects for the first (and in this section, only) outcome.  We define $c=t-1$ to be the number of basic parameters in the univariate setting. Treatment effects  not involving A can be obtained as linear combinations of the basic parameters and are referred to as functional parameters (Lu and Ades, 2006). For example the average treatment effect of treatment E to treatment C, $\delta_1^{CE} = \delta_1^{AE} - \delta_1^{AC}$, is a functional parameter.  We define the $c \times 1$ column vector ${\bm \delta}=(\delta_1^{AB}, \delta_1^{AC}, \cdots, \delta_1^{AZ})^T$  and  design specific $c_d \times c$ design matrices ${\mathbf Z}_{(d)}$. We use the subscript $(d)$ in these design matrices to emphasise that they apply to each individual study of design $d$; we reserve the subscript $d$ for design matrices that describe regression models for all outcome data from this design.    If the $i$th entry of the ${\mathbf Y}_{di}$ are  estimated  treatment effects of treatment J relative to the reference treatment A then the $i$th row of  ${\mathbf Z}_{(d)}$ contains a single nonzero entry: 1 in the $(j-1)$th column, where $j$ is the position of $J$ in the alphabet.  If instead the  $i$th entry of the ${\mathbf Y}_{di}$ are estimated  treatment effects of treatment $J$ relative to treatment  $K$, $K \ne A$,  then the $i$th row of  ${\mathbf Z}_{(d)}$ contains two nonzero entries: 1 in the $(j-1)$th column and -1 in the  $(k-1)$th column.

Our univariate model for network meta-analysis is
\begin{equation}
\label{nine}
{\bm Y}_{di} = {\mathbf Z}_{(d)}  {\bm \delta} + {\bm \Theta}_{di} + {\bm \Omega}_{d} + {\bm \epsilon}_{di}
\end{equation}
where ${\bm \Theta}_{di} \sim N( {\bf 0}, \tau^2_\beta \mathbf{P}_{c_{d}})$, ${\bm \Omega}_{d} \sim N( {\bf 0}, \tau^2_\omega \mathbf{P}_{c_{d}})$, ${\bm \epsilon}_{di} \sim N( {\bf 0}, \mathbf{S}_{di})$, all ${\bm \Theta}_{di}$, ${\bm \Omega}_{di}$ and ${\bm \epsilon}_{di}$ are independent, and  ${\bf P}_{c_{d}}$ is the $c_{d} \times c_{d}$  matrix with ones  on the leading diagonal and halves elsewhere. We refer to $\tau^2_\beta$ and $\tau^2_\omega$ as the between-study variance, and the inconsistency variance, respectively. The term ${\bm \Theta}_{di}$  is a study-by-treatment interaction term that models between-study heterogeneity. The model ${\bm \Theta}_{di} \sim N( {\bf 0}, \tau^2_\beta \mathbf{P}_{c_{d}})$ implies that the heterogeneity variance is the same for all contrasts for every study regardless of whether or not the comparison is relative to the baseline treatment (Lu and Ades, 2004). Other simple choices of $\mathbf{P}_{c_{d}}$, such as allowing the off-diagonal entries to differ from 0.5, violate this symmetry between treatments. For example, in the case  $d=2$ indicating the CDE design, the between-study variances for the CD and CE effects in this study are given by the two diagonal entries of $\tau^2_\beta \mathbf{P}_{c_{d}}$, which are both $\tau^2_\beta$. The between-study variance for the effect of E relative to D is $(-1, 1) \tau^2_\beta \mathbf{P}_{c_{d}} (-1, 1)^T$, which is also $\tau^2_\beta$. The ${\bm \Omega}_{d}$ are design-by-treatment interaction terms that model inconsistency in the network.  The model ${\bm \Omega}_{d} \sim N( {\bf 0}, \tau^2_\omega \mathbf{P}_{c_{d}})$ implies that the  inconsistency variance is the same for all contrasts for every design; other simple choices of $\mathbf{P}_{c_{d}}$ also violate this symmetry.

To describe all estimates from all studies, we  stack the ${\bm Y}_{di}$ from the same design to form  the $n_d \times 1$ column vector ${\bm Y}_d = ({\bm Y}_{d1}^T, \cdots, {\bm Y}^T_{d N_d})^T$, and  we then stack these  ${\bm Y}_d$ to form the $n \times 1$ column vector ${\bm Y} = ({\bm Y}_{1}^T, \cdots, {\bm Y}^T_{D})^T$. Jackson {\it et al.} (2016) then use three further matrices that we also define here because they will be required to describe the estimation procedure that follows.  The matrix ${\bf  M_1}$ is defined as a  $n \times n$ square matrix  where $m_{1ij}=0$ if the $i$th and $j$th entries of ${\bm Y}$, $i,j=1, \cdots n$, are estimates from different studies; otherwise ${m}_{1ii} =1$, and  ${m}_{1ij} =1/2$  for $i \ne j$. The matrix ${\bf  M_2}$ is defined as a  $n \times n$ square matrix  where ${m}_{2ij}=0$  if the $i$th and $j$th entries of ${\bm Y}$, $i,j=1, \cdots n$, are estimates from different designs; otherwise   ${m}_{2ij} =1$ if the $i$th and $j$th entries of ${\bm Y}$ are estimates of the same treatment comparison (for example, treatment A compared to treatment B) and  ${m}_{2ij} =1/2$ if these entries are estimates of different treatment comparisons.   The supplementary materials   show a concrete example showing how these two matrices  are formed. Jackson  {\it et al.} (2016) also define a $n \times c$ univariate design matrix ${\mathbf Z}$,  where if the $i$th entry of ${\mathbf Y}$ is an estimated  treatment effect of treatment J relative to the reference treatment A then the $i$th row of  ${\mathbf Z}$ contains a single nonzero entry: 1 in the $(j-1)$th column, where $j$ is the position of $J$ in the alphabet.  If instead  the $i$th entry ${\mathbf Y}$ is an estimated   treatment effect of treatment $J$ relative to treatment  $K$, $K \ne A$,  then the $i$th row of  ${\mathbf Z}$ contains two nonzero entries: 1 in the $(j-1)$th column and -1 in the  $(k-1)$th column. Defining ${\mathbf S}_d=\mbox{diag}({\mathbf S}_{d1}, \cdots, {\mathbf S}_{d N_d} )$, and then ${\mathbf S}=\mbox{diag}({\mathbf S}_{1}, \cdots, {\mathbf S}_{D} )$, model (\ref{nine}) can be presented for the entire dataset as
\[
{\mathbf Y} \sim N({\mathbf Z} {\bm \delta},   {\tau^2_\beta{\bf M_1}  +  {\tau^2_\omega {\bf M_2}}  }+ {\mathbf S})
\]

\section{A multivariate network meta-analysis model}
We now explain how to extend the univariate model in section 2 to the multivariate setting to handle multiple outcomes. We define $p$ to be the number of outcomes, and so the dimension of the network meta-analysis, so that we now consider the case where $p>1$.   The ${\bm Y}_{di}$ are now $p c_d \times  1$ column vectors, where the ${\bm Y}_{di}$ contain $c_d$ column vectors of length $p$. For example, in a $p=5$ dimensional meta-analysis and continuing with the example where $d=2$ indicates the CDE design, we have $c_2=2$. The ${\bm Y}_{2i}$ are then $10 \times 1$ column vectors where,  taking C as the baseline treatment for this design, the first five entries of the ${\bm Y}_{2i}$ are estimated relative treatment effect of D compared to C and the second five entries are the same estimate of E compared to C. We  define the $pc \times 1$ column vector ${\bm \delta}=(\delta_1^{AB}, \delta_1^{AC}, \cdots, \delta_1^{AZ}, \delta_2^{AB}, \delta_2^{AC}, \cdots, \delta_2^{AZ}, \cdots,  \delta_p^{AB}, \delta_p^{AC}, \cdots, \delta_p^{AZ})^T$, so that this vector contains the basic parameters for each outcome in turn. When $p=1$ the vector ${\bm \delta}$ reduces to its definition in the univariate setting, as given in section 2.

We define ${\bf \Sigma}_\beta$ and ${\bf \Sigma}_\omega$ to be $p \times p$ unstructured covariance matrices that are multivariate generalisations of $\tau^2_\beta$ and $\tau^2_\omega$. These two matrices contain the between-study variances and covariances, and the inconsistency variances and covariances, respectively, for all $p$ outcomes. We refer to ${\bf \Sigma}_\beta$ and ${\bf \Sigma}_\omega$ as the between-study covariance matrix, and the inconsistency covariance matrix, respectively. We continue to treat the within-study covariance matrices ${\bm S}_{di}$ as if fixed and known in analysis but these are now $pc_d \times p c_d$ matrices. The entries of the ${\bm S}_{di}$  matrices that describe the covariance of estimated treatment effects for the same outcome can be obtained as in the univariate setting. However the other entries of ${\bm S}_{di}$, that describe the covariance between treatment effects for different outcomes, are harder to obtain in practice.  A variety of strategies for dealing with this difficulty have been proposed (Jackson {\it et al}., 2011; Wei and Higgins, 2013).

\subsection{The proposed multivariate model for network meta-analysis}
In the multivariate setting, to allow correlations between estimated treatment effects for different outcomes, both within studies and designs,  we propose that model (\ref{nine}) is generalised to
\begin{equation}
\label{new_label}
{\bm Y}_{di} =   {\bm X}_{(d)} {\bm \delta} + {\bm \Theta}_{di} + {\bm \Omega}_{d} + {\bm \epsilon}_{di}
\end{equation}
where  ${\bm X}_{(d)} =(({\bm I}_p \otimes {\bm Z}_{(d)1})^T, \cdots, ({\bm I}_p \otimes {\bm Z}_{(d) c_d})^T)^T $, ${\bm Z}_{(d)i}$ is the $i$th row of ${\bm Z}_{(d)}$,
$
 {\bm \Theta}_{di} \sim N( {\bf 0}, \mathbf{P}_{c_{d}} \otimes {\bf \Sigma}_\beta)
 $,
 $
 {\bm \Omega}_{d} \sim N( {\bf 0},  \mathbf{P}_{c_{d}} \otimes {\bf \Sigma}_\omega)
 $
 and
 $
 {\bm \epsilon}_{di} \sim N( {\bf 0}, {\bm S}_{di})$, where all ${\bm \Theta}_{di}$, ${\bm \Omega}_{d}$ and ${\bm \epsilon}_{di}$ are independent, and $\otimes$ is the Kronecker product.  The random ${\bm \Theta}_{di}$ and ${\bm \Omega}_{d}$ continue to model between-study heterogeneity, and inconsistency, respectively. Recalling that ${\bm \delta}$ contains the basic parameters for each outcome in turn, the  design matrices ${\bm X}_{(d)}$ provide the correct linear combinations of basic parameters to describe the mean of all estimated treatment effects in ${\bm Y}_{di}$. Model (\ref{new_label}) reduces to model (\ref{nine}) in one dimension. The definition of $\mathbf{P}_{c_{d}}$ means that ${\bf \Sigma}_\beta$ and ${\bf \Sigma}_\omega$ are the between-study covariance matrix, and inconsistency covariance matrix, for all contrasts. We continue define ${\bm Y}$ as in the univariate setting, where ${\bm Y}$ contains $n$ column vectors of estimated treatment effects that are of length $p$, so that ${\bm Y}$ is a $np \times 1$ column vector in the multivariate setting. We define the multivariate $np \times pc $ design matrix ${\mathbf X}= (({\mathbf I}_p \otimes {\mathbf Z}_{1})^T, \cdots, ({\mathbf I}_p \otimes {\mathbf Z}_{n})^T)^T$, where ${\mathbf Z}_{i}$ is the $i$th row of ${\mathbf Z}$. Model (\ref{new_label}) can be presented for the entire dataset as
\begin{equation}
\label{nine1}
{\mathbf Y} \sim N({\mathbf X} {\bm \delta},   {{\bf M_1} \otimes {\bf \Sigma}_{\beta} +  {{\bf M_2}}\otimes {\bf \Sigma}_{\omega} }+ {\mathbf S})
\end{equation}
 where we continue to define ${\mathbf S}$ as in the univariate case. Matrices ${\bf M_1}$ and ${\bf M_2}$ are the same as in the univariate setting, and so continue to be $n \times n$ matrices.
 Model (\ref{nine1}) is a linear mixed model for network meta-analysis and is conceptually similar to other models of this type (Piepho {\it et al.}, 2012). If ${\bf \Sigma}_{\omega}={\bf 0}$ then all ${\bm \Omega}_{d}= {\bm 0}$ and there is no inconsistency; we refer to this reduced model as the `consistent model'. If both ${\bf \Sigma}_{\beta}={\bf 0}$ and ${\bf \Sigma}_{\omega}={\bf 0}$ then all studies estimate the same effects to within-study sampling error and we refer to this model as the `common-effect and consistent model'.

 Missing data (unobserved entries of ${\mathbf Y}$) are common in applications as not all studies may provide data for all outcomes and contrasts. When there are missing outcome data, the model for the observed data is the marginal model for the observed data implied by (\ref{nine1}), where any rows of ${\mathbf Y}$ that contain missing values are discarded. We will use a non-likelihood based approach for making inferences and so assume any data are missing completely at random (Seaman {\it et al.}, 2013). We define the diagonal $np \times np$ missing indicator matrix ${\mathbf R}$, where ${\mathbf R}_{ii}=1$ if ${\mathbf Y}_i$ is observed, ${\mathbf R}_{ii}=0$ if ${\mathbf Y}_i$ is missing, and ${\mathbf R}_{ij}=0$ if $i \ne j$.

\section{Multivariate estimation: a new method of moments}
 Our estimation procedure is motivated by the univariate method proposed by DerSimonian and Laird (1986). This  was developed in the much simpler setting where each study provides a single estimate $Y_i$, and where the random-effects model $Y_i \sim N(\delta, \tau^2+S_i)$ is assumed.  This estimation method for $\tau^2$ uses the $Q$ statistic, where $Q=\sum S_i^{-1}(Y_i -\hat{\delta})^2$ and   $\hat{\delta}  = \sum S_i^{-1} Y_i/\sum S_i^{-1}$  is the pooled estimate under the common-effect model ($\tau^2=0$).

Now consider an alternative representation of this $Q$ statistic. Taking  ${\bf Y} = (Y_1, \cdots, Y_n)^T$, ${\bf S} = \mbox{diag}(S_1, \cdots, S_n)$ and ${ \bf W}={\bf S}^{-1}$ means that $Q=\mbox{tr}({\mathbf W} ({\mathbf Y}-\hat{{\mathbf Y}})  ({\mathbf Y}-\hat{{\mathbf Y}})^{T})$, where $\hat{{\mathbf Y}}$ is obtained under the common-effect model. To obtain a $p \times p$ matrix generalisation of $Q$ for multivariate analyses, we replace the trace operator with the block trace operator in this expression (Jackson {\it et al}., 2013). The block trace operator is a generalisation of the trace  that sums over all  $n$ of the $p \times p$ matrices along the main block diagonal of an $np \times np$ matrix. This produces a $p \times p$ matrix.  In the absence of missing data we can  write our multivariate generalisation of the $Q$ statistic, $\mbox{btr}({\mathbf W} ({\mathbf Y}-\hat{{\mathbf Y}})  ({\mathbf Y}-\hat{{\mathbf Y}})^{T})$, as a weighted sum of outer products of $p \times 1$ vectors of residuals under the common-effect and consistent model. Hence the distribution of $\mbox{btr}({\mathbf W} ({\mathbf Y}-\hat{{\mathbf Y}})  ({\mathbf Y}-\hat{{\mathbf Y}})^{T})$ depends directly on the magnitudes of unknown variance components.

\subsection{A Q matrix for multivariate network meta-analysis}
We define a within-study precision matrix ${\mathbf W}$ corresponding to ${\mathbf S}$. If there are no missing outcome data in ${\mathbf Y}$  then we define ${\mathbf W} = {\mathbf S}^{-1}$, where ${\mathbf S}$ is taken from model (\ref{nine1}). If there are missing data in ${\mathbf Y}$ then the entries of ${\mathbf W}$ that correspond to observed data are obtained as the inverse of the corresponding entries of the within-study covariance matrix of reduced dimension (equal to that of the observed data) and the other entries of ${\mathbf W}$ are set to zero. For example, consider the case where  ${\mathbf Y}$ is a $6 \times 1$ vector but only the second and fifth entries are observed; this corresponds to much less outcome data than would be used in practice but provides an especially simple example. Then we define ${\mathbf S}_{r}$, where the subscript $r$ indicates a dimension reduction, as a $2 \times 2$ matrix whose entries are the within variances and covariances of the two observed entries of ${\mathbf Y}$. The $6 \times 6$ precision matrix ${\mathbf W}$ then has all zero entries in the first, third, fourth and sixth rows and columns. However the remaining entries of ${\mathbf W}$ are the entries of the $2 \times 2$ matrix ${\mathbf S}_{r}^{-1}$, so that ${\mathbf W}_{22} = ({\mathbf S}_{r}^{-1})_{11}$, ${\mathbf W}_{25} = ({\mathbf S}_{r}^{-1})_{12}$, ${\mathbf W}_{52} = ({\mathbf S}_{r}^{-1})_{21}$, and ${\mathbf W}_{55} = ({\mathbf S}_{r}^{-1})_{22}$. We define $\hat{{\mathbf Y}}$ to be the fitted value of  ${{\mathbf Y}}$ under the common-effect and consistent model (${\bf \Sigma}_{\beta} = {\bf \Sigma}_{\omega}= {\bf{0}}$), so that $ \hat{{\mathbf Y}} =  {\mathbf H} {\mathbf Y}$ where ${\mathbf H} = {\mathbf X} ({\mathbf X}^T {\mathbf W} {\mathbf X})^{-1} {\mathbf X}^{T} {\mathbf W}$. We also define an asymmetric $np \times np$  matrix (Jackson {\it et al.}, 2013)
\begin{equation}
\label{eq5} {\mathbf Q}= {\mathbf W} \{{\mathbf R} ({\mathbf Y}-\hat{{\mathbf Y}})\}  \{{\mathbf R}({\mathbf Y}-\hat{{\mathbf Y}})\}^{T}   =  {\mathbf W} ({\mathbf Y}-\hat{{\mathbf Y}})  ({\mathbf Y}-\hat{{\mathbf Y}})^{T} {\mathbf R}
\end{equation}
Our definitions of ${\mathbf W}$ and ${\mathbf R}$ mean that ${\mathbf W} {\mathbf R}={\mathbf W}$, which results in the simplified version of ${\mathbf Q}$ in (\ref{eq5}). From the first form given in (\ref{eq5}), we have that the residuals ${\mathbf Y}-\hat{\mathbf Y}$ are pre-multiplied by ${\mathbf R}$, so that any residuals that correspond to missing outcome data do not contribute to  ${\mathbf Q}$. Furthermore missing outcome data do not contribute to  $\hat{\mathbf Y}$ because they have no weight under the common-effect and consistent model. Hence we can impute missing outcome data with any finite value without changing the value of ${\mathbf Q}$. This is merely a convenient way to handle missing data numerically and  has no implications for the statistical modelling.

  \subsection{Design specific Q matrices for multivariate network meta-analysis}
  In order to identify the full model, we will require design-specific versions of  ${\mathbf Q}$ that only use data from a particular design.
  As in the univariate setting, we stack the outcome data from design $d$ to form the  vector ${\mathbf Y}_d = ({\mathbf Y}^T_{d1}, \cdots, {\mathbf Y}^T_{dN_d})^T$. In the multivariate setting the vector ${\mathbf Y}_d$ contains $n_d$ estimated effects each of length $p$, so that ${\mathbf Y}_d$ is now a $p n_d \times 1$ column vector. We  define the design specific  $n_d \times n_d$ matrix ${\bf M_1^d}$, where ${m}^d_{1ij}=0$  if the $i$th and $j$th estimated effect (of length $p$)  in ${\mathbf Y}_d$, $i,j=1, \cdots, n_d$, are from separate studies; otherwise ${m}^d_{1ii} =1$ and  ${m}^d_{1ij} =1/2$  for $i \ne j$.   We define the $ p n_d \times p c_d $  design matrix ${\mathbf X}_{d}$ which is obtained by stacking identity matrices of dimension $p c_d$, where we include one such identity matrix for each study of design $d$. Hence ${\mathbf X}_{d} = {\bf 1}_{N_d} \otimes {\mathbf I}_{p c_{d}}$, where ${\bf 1}_{N_d}$ is the $N_d \times 1$ column vector where every entry is one. We also define the $p c_d \times 1$ column vector ${\bm \beta}_{d} = {\bm X}_{(d)} {\bm \delta} + {\bm \Omega}_{d}$.

  An identifiable  design-specific marginal model for outcome data from design $d$ only, that is implied by model (\ref{new_label}), is
\begin{equation}
\label{nined}
{\mathbf Y}_d \sim N({\mathbf X}_{d} {\bm \beta}_{d},   {{\bf M_1^d} \otimes {\bf \Sigma}_{\beta}}+ {\mathbf S}_d )
\end{equation}
where  ${\mathbf S}_d=\mbox{diag}({\mathbf S}_{d1}, \cdots,{\mathbf S}_{d N_d})$. We can also calculate design specific versions of (\ref{eq5})  where we calculate all quantities, including the fitted values, using just the data from studies of design $d$. We define these $p n_d  \times p n_d$ design specific matrices as
  \begin{equation}
\label{cactus}
  {\mathbf Q}_d  = {\mathbf W}_d ({\mathbf Y}_d-\hat{{\mathbf Y}}_d)  ({\mathbf Y}_d-\hat{{\mathbf Y}}_d)^{T} {\mathbf R}_d
 \end{equation}
  where ${\mathbf W}_d$, ${\mathbf R}_d$ and $\hat{{\mathbf Y}}_d$ in (\ref{cactus}) are defined in the same way as ${\mathbf W}$, ${\mathbf R}$ and $\hat{{\mathbf Y}}$ in (\ref{eq5}) but where only data from design $d$ are used. Hence ${\mathbf R}_d$ and ${\mathbf W}_d$ are the missing indicator matrix, and the within-study precision matrix, of ${\mathbf Y}_d$, respectively. We compute $\hat{{\mathbf Y}}_d = {\mathbf H}_d {\mathbf Y}_d$ where ${\mathbf H}_d = {\mathbf X}_d ({\mathbf X}_d^T {\mathbf W}_d {\mathbf X}_d)^{-1} {\mathbf X}_d^{T} {\mathbf W}_d$. When computing ${\mathbf H}_d$ we take the matrix inverse to be the Moore-Penrose pseudoinverse. This is so that any  design-specific regression corresponding to this hat matrix that is not fully identifiable (due to missing outcome data) can still contribute to the estimation. We use  model (\ref{nined}) to derive the properties of  ${\mathbf Q}_d$ in equation (\ref{cactus}).

  \subsection{The estimating equations}
 We  base our estimation on the two $p \times p$ matrices
 $\mbox{btr} ({\mathbf Q}) $
  and $  \sum\limits_{d=1}^D\mbox{btr} ({\mathbf Q}_d)$, where ${\mathbf Q}$ and ${\mathbf Q}_d$ are given in (\ref{eq5}) and (\ref{cactus}), respectively. Specifically, we match these quantities to their expectations to estimate the unknown variance parameters using the method of moments.

 \subsubsection{Evaluating $\mbox{E}[\mbox{btr} ({\mathbf Q})] $ and deriving the first estimating equation}
We define ${\mathbf A} = ({\mathbf I}_{np}-{\mathbf H})^T {\mathbf W}$ and ${\mathbf B} = ({\mathbf I}_{np} - {\mathbf H})^{T} {\mathbf R}$, which are known $np \times np$ matrices. We also divide the matrices ${\mathbf A}$ and ${\mathbf B}$ into $n^2$ blocks of  $p \times p$ matrices, and write ${\mathbf A}_{i,j}$ and ${\mathbf B}_{i,j}$, $i,j=1, \cdots n$, to mean the $i$th by $j$th blocks of  ${\mathbf A}$ and ${\mathbf B}$ respectively. Hence ${\mathbf A}_{i,j}$ and ${\mathbf B}_{i,j}$ are both $p \times p$ matrices.  In the supplementary materials we show that
\[
\mbox{E}[\mbox{btr}({\mathbf Q})]=
\sum\limits_{i=1}^{n}\sum\limits_{j=1}^{n}\sum\limits_{k=1}^{n} m_{1i j} {\mathbf A}_{k,i} {\bf \Sigma}_\beta {\mathbf B}_{j,k}+
\]
\[
\sum\limits_{i=1}^{n}\sum\limits_{j=1}^{n}\sum\limits_{k=1}^{n} m_{2i j} {\mathbf A}_{k,i} {\bf \Sigma}_\omega {\mathbf B}_{j,k} + \mbox{btr}({\mathbf B}).
\]
We apply the $\mbox{vec}(\cdot)$ operator to both sides of the previous equation and use the identity $\mbox{vec}({\mathbf A}{\mathbf X}{\mathbf B}) = ({\mathbf B}^T \otimes {\mathbf A})\mbox{vec}({\mathbf X})$ (see Henderson and Searle, 1981), to obtain
\begin{equation}
\label{eq1}
\mbox{vec}(\mbox{E}[\mbox{btr}({\mathbf Q})]) = {\mathbf C} \mbox{vec}({\bf \Sigma}_\beta) + {\mathbf D} \mbox{vec}({\bf \Sigma}_\omega) + {\mathbf E}
\end{equation}
where
\[
{\mathbf C} =   \sum\limits_{i=1}^{n}\sum\limits_{j=1}^{n}\sum\limits_{k=1}^{n} m_{1i j} {\mathbf B}_{j,k}^T \otimes {\mathbf A}_{k,i}
\]
\[
{\mathbf D} =  \sum\limits_{i=1}^{n}\sum\limits_{j=1}^{n}\sum\limits_{k=1}^{n} m_{2i j} {\mathbf B}_{j,k}^T \otimes {\mathbf A}_{k,i}
\]
and
\[
{\mathbf E} = \mbox{vec}(\mbox{btr}({\bf B})).
\]
Upon substituting $\mbox{E}[\mbox{btr}({\mathbf Q})] = \mbox{btr}({\mathbf Q})$,  ${\bf \Sigma}_\beta = {\hat{\bf \Sigma}}_\beta$ and ${\bf \Sigma}_\omega=\hat{{\bf \Sigma}}_\omega$ in equation (\ref{eq1}), the method of moments gives one estimating equation in the vectorised form of two unknown covariance matrices.
\subsubsection{Evaluating $\mbox{E}[\mbox{btr} ({\mathbf Q}_d)] $ and deriving the second estimating equation}
Model (\ref{nined}) depends upon one unknown covariance matrix, ${\bf \Sigma}_{\beta}$. The intuition is that, upon using all $D$ of the ${\mathbf Q}_d$ matrices in (\ref{cactus}) and the method of moments to estimate ${\bf \Sigma}_{\beta}$,  we will then be able to  estimate the other unknown covariance matrix ${\bf \Sigma}_\omega$ using the first estimating equation.
We define design specific $
{\mathbf A}_d = ({\mathbf I}_{p n_d}-{\mathbf H}_d)^T {\mathbf W}_d
$
and
$
{\mathbf B}_d = ({\mathbf I}_{p n_d} - {\mathbf H}_d)^{T} {\mathbf R}_d
$
, where ${\mathbf A}_d$ and ${\mathbf B}_d$ are known $p n_d \times p n_d$ matrices. We also divide the matrices ${\mathbf A}$ and ${\mathbf B}$ into $n_d^2$ blocks of  $p \times p$ matrices, and write ${\mathbf A}_{d, i,j}$ and ${\mathbf B}_{d, i,j}$, $i,j=1, \cdots, n_d$, to mean the $i$th by $j$th blocks of  ${\mathbf A}_d$ and ${\mathbf B}_d$ respectively. In the supplementary materials we show that
\begin{equation}
\label{eq3}
\mbox{vec}\left(\mbox{E}[\sum\limits_{d=1}^D \mbox{btr}({\mathbf Q}_d)]\right) = \left(\sum\limits_{d=1}^D {\mathbf C}_d\right) \mbox{vec}({\bf \Sigma}_\beta)  + \sum\limits_{d=1}^D {\mathbf E}_d
\end{equation}
where
\[
{\mathbf C}_d = \sum\limits_{i=1}^{n_d}\sum\limits_{j=1}^{n_d}\sum\limits_{k=1}^{n_d} m^d_{1 ij} {\mathbf B}_{d, j,k}^T \otimes {\mathbf A}_{d, k,i}
\]
and
\[
{\mathbf E}_d = \mbox{vec}(\mbox{btr}({\bf B}_d)).
\]
Upon substituting $\mbox{E}[\sum\limits_{d=1}^D \mbox{btr}({\mathbf Q}_d)]=\sum\limits_{d=1}^D \mbox{btr}({\mathbf Q}_d)$ and ${\bf \Sigma}_\beta = \hat{{\bf \Sigma}}_\beta$ in (\ref{eq3}), we obtain a second estimating equation from the method of moments.

 \subsection{Solving the estimating equations and performing inference}
 We solve the estimating equation resulting from (\ref{eq3}) for $\mbox{vec}({\bf \hat{\Sigma}}_\beta)$ and substitute this estimate into the estimating equation resulting from (\ref{eq1}) and solve for  $\mbox{vec}({\bf \hat{\Sigma}}_\omega)$.
\subsubsection{Estimating ${\bf \Sigma}_\beta$ under the consistent model}

 Some applied analysts may prefer to  assume the consistent model (${\bf \Sigma}_\omega  = {\bf 0})$. As in the univariate case (Jackson  {\it et al.}, 2016), we have two possible ways of estimating ${\bf \Sigma}_\beta$ under the consistent model: we can use the estimating equation resulting from (\ref{eq1}) with ${\bf \Sigma}_\omega = {\bf 0}$ or the estimating equation resulting from  (\ref{eq3}) as in the full model. Also as in the univariate case, we suggest the former option because it uses the information made by assuming consistency when estimating ${\bf \Sigma}_\beta$. However this first option  is valid only under the consistent model.
\subsubsection{`Truncating' the estimates of the unknown covariance matrices so that they are symmetric and positive semi-definite}
As in the univariate case, there is the problem that the point estimates of the two unknown covariance matrices are not necessarily positive semi-definite. The method of moments does not even initially enforce the constraint that the point estimates of the unknown covariance matrices are symmetrical (Chen {\it et al.}, 2012; Jackson {\it et al.}, 2013). We produce symmetric estimators corresponding to an estimated covariance matrix of $\hat{{\bf \Sigma}}$ as $(\hat{{\bf \Sigma}^T} + {\bf \hat{\Sigma}})/2$  (Chen {\it et al.}, 2012; Jackson {\it et al.}, 2013). This also corresponds to taking the average of estimates that result from our ${\bf Q}$ and ${\bf Q}_d$ matrices and their transposes (Jackson  {\it et al.}, 2013). We then write these symmetric estimators in terms of their spectral decomposition (Chen {\it et al.}, 2012; Jackson {\it et al.}, 2013) and truncate any negative eigenvalues to zero to provide the final symmetric positive semi-definite estimated covariance matrices. Specifically, we define the truncated estimate  corresponding to the symmetrical $\hat{{\mathbf{\Sigma}}}$ as
$
\hat{{\mathbf{\Sigma}}}^{+} = \sum\limits_{i=1}^{p} \mbox{max}(0, \lambda_i)
{\mathbf e}_i {\mathbf e}_i^T,
$
where $\lambda_i$ is the $i$th eigenvalue of the symmetric  $\hat{\mathbf{\Sigma}}$ and ${\mathbf e}_i$ is the corresponding normalised eigenvector.

\subsubsection{Inference for ${\bm \delta}$}
Inference for ${\bm \delta}$ then proceeds as a weighted regression where all weights are treated as fixed and known. Writing $\hat{\bf{V}}$ as the estimated variance of ${\bf Y}$ in (\ref{nine1}), in the absence of missing outcome data we have
$
\hat{{\bm \delta}}= ({\bm X}^T \hat{\bf{V}}^{-1} {\bm X})^{-1} {\bm X}^T \hat{\bf{V}}^{-1} {\bm Y}
$
where
$
\mbox{Var}(\hat{{\bm \delta}}) = ({\bm X}^T \hat{\bf{V}}^{-1} {\bm X})^{-1}.
$
In the presence of missing data we can, under our missing completely at random assumption,  apply these standard formulae for weighted regression to the observed outcomes. Alternatively and equivalently,  we can impute the missing outcome data in ${\bf Y}$ with an arbitrary value and replace  $\hat{\bf V}^{-1}$ with the precision matrix corresponding to $\hat{\bf V}$,  calculated in the way explained for ${\bm S}$ in section 4.1 (Jackson {\it et al.}, 2011).
Approximate confidence intervals and hypothesis tests for all basic parameters for all outcomes then immediately follow by taking $\hat{{\bm \delta}}$ to be approximately normally distributed. Inferences for functional parameters follow by taking appropriate linear combinations of $\hat{{\bm \delta}}$.

\subsection{Special cases of the estimation procedure}
In the supplementary materials we show that the proposed method reduces to two previous methods in special cases. If all studies are two arm studies  and consistency is assumed then the proposed method reduces to the matrix based method for multivariate meta-regression  (Jackson {\it et al.}, 2013). The proposed multivariate method  reduces to the univariate DerSimonian and Laird method for network meta-analysis (Jackson {\it et al.}, 2016) when $p=1$.
\subsection{Model identification}

If the necessary standard matrix inversions resulting from the estimating equations from (\ref{eq1}) and (\ref{eq3}) cannot be performed then both unknown variance components cannot be identified using the proposed method.  A minimum requirement for any multivariate modelling is that the common-effect and consistent model must be identifiable. This means that there must be some information (direct or indirect) about each basic parameter for all outcomes. Two or more studies of the same design must provide data for all possible pairs of outcomes to identify ${\bf \Sigma}_{\beta}$. Two or more studies of different designs must provide data for all possible pairs of outcomes to identify  ${\bf \Sigma}_{\omega}$. If these conditions are satisfied then the model will be identifiable. In situations where our model is not  identifiable  we suggest that simpler models should be considered instead. Possible strategies for this include considering models of lower dimension or the consistent model. In practice it is highly desirable to have more than the minimum amount of replication required, both within and between designs, so that the model is well identified. We make some pragmatic decisions in the next section for our example to provide sufficient replication within designs, in order to estimate ${\bf \Sigma}_{\beta}$ with reasonable precision.

\section{Example}
The methodology developed in this paper is now applied to an illustrative example in relapsing remitting multiple sclerosis (RRMS). Multiple sclerosis (MS) is an inflammatory disease of the brain and spinal cord and RRMS is a common type of MS.  The effectiveness of a new treatment is typically measured to assess its impact on relapse rate and odds of disease progression. Magnetic Resonance Imaging (MRI)  allows measurement of the number of new or enlarging lesions in the brain. Three outcomes are included in our analyses, so that $p=3$ in the full three dimensional network meta-analysis. These three outcomes are: (1) the log rate ratio of new or enlarging MRI lesions; (2) the log annualised relapse rate ratio; and (3) log disability progression odds ratio. Relapse is defined as appearance of new, worsening or recurrence of neurological symptoms that can be attributable to MS, accompanied by an increase of  a score on the Expanded Disability Status Scale (EDSS) and also functional-systems score(s), lasting at least 24 hours, preceded by neurologic stability for at least 30 days. Disability progression is defined as an increase in EDSS score that was sustained for 12 weeks, with an absence of relapse at the time of assessment. Negative basic parameters indicate that treatments B-F are beneficial compared to treatment A throughout.

Data in this illustrative example were obtained from ten randomised controlled trials of six treatment options (coded in the network data as treatments A to F); placebo (A), interferon beta-1b (B), interferon beta-1a (C), glatiramer (D), and two doses of fingolimod; 0.5mg (E) and 1.25mg (F). Three of the fingolimod trials were three-arm (two doses and a control) and are included as three-arm studies. Three trials of interferon beta (one 1a and two 1b) were three-arm (also two doses and a control), and these were included as separate two-arm trials (each dose against the control, with the number of participants in each control arm halved). This  ignores the differences in doses of interferon beta and was a pragmatic decision to help provide an identifiable network. Briefly, in this example there is very little replication within designs, so that identifying ${\bf \Sigma}_\beta$ well is very difficult without making pragmatic decisions such as this.  Sormani {\it et al.} (2010) also treat these particular studies as two separate studies in this way, which helps them to identify their meta-regression models. Treating these three studies as separate two-arm trials means that the data are analysed as being from thirteen studies and a summary of the resulting data structure is shown in Table \ref{tab-ms-data}. There are eight different designs in Table \ref{tab-ms-data} and so there is relatively little replication within designs, even when including three of the three-arm studies as separate two arm studies. See  Bujkiewicz {\it et al}. (2016) for further details of these data. Figure 1 provides network diagrams that show the number of comparisons between each pair of treatments on the edges. In these diagrams the three arm studies (Table \ref{tab-ms-data}) are taken to contribute three comparisons, for example the CEF study contributes CE, CF and EF comparisons. Two estimates of treatment effect from this study contribute to analyses however because C is taken as the baseline; the study's estimated EF treatment effect  contains no additional information once its CE and CF contrasts are included in the analysis.

\begin{figure}
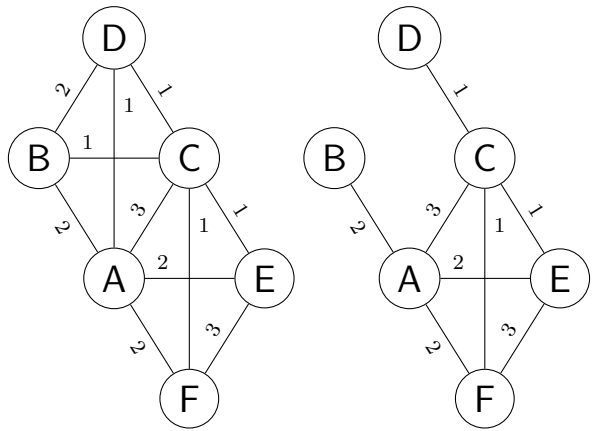

\caption{\label{netdiag}Network diagram for RRMS dataset. A -- placebo, B -- interferon beta-1b, C -- interferon beta-1a, D -- glatiramer, E -- fingolimod 0.5mg, F -- fingolimod 1.25mg. Left-hand-side network corresponds to studies reporting the log annualised relapse rate ratio and log disability
progression odds ratio ($y_2$ and $y_3$) for which data are complete. The right-hand-side network corresponds to studies reporting the log rate ratio of new or enlarging MRI lesions ($y_1$ which is not reported in four studies). The numbers shown on the network edges are the number of direct comparisons of each pair of treatments; the absence of an edge indicates that there is no direct comparison. Three of the thirteen studies are three arm trials which are each taken to provide three direct comparisons (a direct comparison between each treatment pair). Hence there are 19 direct comparisons in the left-hand-side network where there is no missing data. }
\vspace{0.25cm}
\tikz
\tikzset{my state/.style={circle,draw}}
[inner sep=3mm]
\path node at ( 0,1.6) [my state] (f) {\Large{$\sf{D}$}}
node at ( -1,0) [my state] (s1) {\Large{$\sf{B}$}}
node at ( 1,0) [my state] (s2) {\Large{$\sf{C}$}}
node at ( 0,-1.6) [my state] (s3) {\Large{$\sf{A}$}}
node at ( 2,-1.6) [my state] (s4) {\Large{$\sf{E}$}}
node at ( 1,-3.2) [my state] (s5) {\Large{$\sf{F}$}}
(s1) edge node[above,sloped] {\footnotesize 2} (f)
(s2) edge node[above,sloped] {\footnotesize 1} (f)
(s1) edge node[above,pos=0.2] {\footnotesize 1} (s2)
(s2) edge node[above,sloped] {\footnotesize 3} (s3)
(s1) edge node[below,sloped] {\footnotesize 2} (s3)
(s3) edge node[above,pos=0.2] {\footnotesize 2} (s4)
(s2) edge node[above,sloped] {\footnotesize 1} (s4)
(s3) edge node[pos=0.8,right] {\footnotesize 1} (f)
(s5) edge node[pos=0.8,right] {\footnotesize 1} (s2)
(s5) edge node[below,sloped] {\footnotesize 2} (s3)
(s4) edge node[above,sloped] {\footnotesize 3} (s5);
\tikz
\tikzset{my state/.style={circle,draw}}
[inner sep=3mm]
\path node at ( 0,1.6) [my state] (f) {\Large{$\sf{D}$}}
node at ( -1,0) [my state] (s1) {\Large{$\sf{B}$}}
node at ( 1,0) [my state] (s2) {\Large{$\sf{C}$}}
node at ( 0,-1.6) [my state] (s3) {\Large{$\sf{A}$}}
node at ( 2,-1.6) [my state] (s4) {\Large{$\sf{E}$}}
node at ( 1,-3.2) [my state] (s5) {\Large{$\sf{F}$}}
(s2) edge node[above,sloped] {\footnotesize 1} (f)
(s2) edge node[above,sloped] {\footnotesize 3} (s3)
(s1) edge node[below,sloped] {\footnotesize 2} (s3)
(s3) edge node[above,pos=0.2] {\footnotesize 2} (s4)
(s2) edge node[above,sloped] {\footnotesize 1} (s4)
(s5) edge node[pos=0.8,right] {\footnotesize 1} (s2)
(s5) edge node[below,sloped] {\footnotesize 2} (s3)
(s4) edge node[above,sloped] {\footnotesize 3} (s5);
\end{figure}

\begin{table}[h]
\caption{\label{tab-ms-data}Summary of the relapsing remitting multiple sclerosis dataset.}
\begin{tabular}{lll}
\hline
\textbf{Study}  & Design & Outcomes  \\
\hline
IFNB SG (1)	&	AB	&	All three outcomes measured\\
IFNB SG (2)	&	AB	&	All three outcomes measured\\
Jacobs/Simon	&	AC	&	All three outcomes measured\\
PRISMS (1)	&	AC	&	All three outcomes measured\\
PRISMS (2)	&	AC	&	All three outcomes measured\\
Johnson	&	AD	&	Relapse rate and disability progression only\\
Durelli	&	BC	&	Relapse rate and disability progression only		\\
O'Connor (1)	&	BD	&	Relapse rate and disability progression only \\
O'Connor (2)	&	BD	&	Relapse rate and disability progression only	\\
Mikol	&	CD	&	All three outcomes measured\\
FREEDOMS 1 	&	AEF	& All three outcomes measured	\\
FREEDOMS 2 	&	AEF	&	All three outcomes measured\\
TRANSFORMS 	&	CEF	&	All three outcomes measured \\
\hline
\end{tabular}
\end{table}

\begin{sidewaystable}
\caption{\label{tab-ms-results1}Treatment effect estimates of each treatment relative to the reference treatment A (placebo).}
\begin{tabular}{lccccc}
\hline
model & \multicolumn{5}{c}{estimate (se)}  \\
      & AB & AC & AD & AE & AF  \\ \hline
\emph{MRI} ($y_1$) & & & & &  \\ \hline
univariate ($y_1$)          & -0.95 (0.39) & -1.00 (0.21) & -0.68 (0.50) & -1.38 (0.26) & -1.52 (0.26) \\
bivariate ($y_1, y_2$)      & -0.94 (0.39) & -1.00 (0.21) & -0.68 (0.50) & -1.39 (0.26) & -1.53 (0.26) \\
bivariate ($y_1, y_3$)      & -0.96 (0.39) & -0.98 (0.22) & -0.66 (0.50) & -1.38 (0.26) & -1.51 (0.26) \\
trivariate ($y_1, y_2, y_3$)& -0.96 (0.39) & -0.97 (0.22) & -0.67 (0.50) & -1.38 (0.26) & -1.51 (0.26) \\
\hline
\emph{Relapse rate} ($y_2$)& & & & &  \\ \hline
univariate ($y_2$)          & -0.35 (0.10) & -0.25 (0.09) & -0.34 (0.11) & -0.81 (0.12) & -0.78 (0.12) \\
bivariate ($y_1, y_2$)      & -0.35 (0.10) & -0.25 (0.09) & -0.34 (0.11) & -0.81 (0.12) & -0.78 (0.12) \\
bivariate ($y_2, y_3$)      & -0.36 (0.11) & -0.23 (0.10) & -0.33 (0.12) & -0.80 (0.13) & -0.77 (0.13) \\
trivariate ($y_1, y_2, y_3$)& -0.36 (0.11) & -0.23 (0.10) & -0.33 (0.12) & -0.80 (0.13) & -0.77 (0.13) \\
\hline
\emph{Disability progression} ($y_3$) & & & & &  \\ \hline
univariate ($y_3$)          & -0.46 (0.25) & -0.11 (0.21) & -0.42 (0.25) & -0.33 (0.25) & -0.37 (0.24) \\
bivariate ($y_2, y_3$)      & -0.47 (0.25) & -0.10 (0.21) & -0.43 (0.25) & -0.37 (0.25) & -0.37 (0.25) \\
bivariate ($y_1, y_3$)      & -0.46 (0.25) & -0.11 (0.21) & -0.42 (0.25) & -0.34 (0.25) & -0.38 (0.25) \\
trivariate ($y_1, y_2, y_3$)& -0.47 (0.25) & -0.10 (0.21) & -0.43 (0.25) & -0.37 (0.25) & -0.37 (0.25) \\ \hline
\end{tabular}
\end{sidewaystable}

\begin{table}
\caption{\label{tab-ms-results2}Inconsistency and heterogeneity covariance matrices estimates.}
\begin{tabular}{lccccccccc}
\hline
model    & $\Sigma_{\omega 11}$  & $\Sigma_{\omega 12}$ & $\Sigma_{\omega 13}$ & $\Sigma_{\omega 22}$ & $\Sigma_{\omega 23}$ & $\Sigma_{\omega 33}$ &&& \\ \hline
univariate ($y_1$)          & 0.0000  & & & & & &&&\\
univariate ($y_2$)          &  & & &0.0115 & & &&&\\
univariate ($y_3$)          & & & & & & 0.0713 &&&\\
bivariate ($y_1, y_2$)      & 0.0002 & 0.0017 & & 0.0125 & & &&& \\
bivariate ($y_1, y_3$)      & 0.0018 & & 0.0116 & & & 0.0741 &&&\\
bivariate ($y_2, y_3$)      & & & &0.0161 & 0.0344 & 0.0735 &&&\\
trivariate ($y_1, y_2, y_3$)& 0.0027& 0.0066& 0.0143 & 0.0161 & 0.0349& 0.0756 &&& \\
\hline
    & $\Sigma_{ \beta 11}$  & $\Sigma_{ \beta 12} $ & $\Sigma_{\beta 13}$ & $\Sigma_{\beta 22 }$ & $\Sigma_{\beta 23} $ & $\Sigma_{\beta 33}$ \\ \hline
univariate ($y_1$)          &  0.1508  & & & & & &&&\\
univariate ($y_2$)          &   & & &0.0043 & & &&&\\
univariate ($y_3$)          & & & & & & 0.0000 &&&\\
bivariate ($y_1, y_2$)      & 0.1523 & -0.0110 & & 0.0047 & & && & \\
bivariate ($y_1, y_3$)      & 0.1526 & & -0.0191 & & & 0.0024 & &  & \\
bivariate ($y_2, y_3$)      & & & &0.0061 & 0.0015 & 0.0004 & & &\\
trivariate ($y_1, y_2, y_3$)& 0.1538 & -0.0116 & -0.0195 & 0.0059 & 0.0024 & 0.0027 & & & \\ \hline
\end{tabular}
\end{table}

Table \ref{tab-ms-results1} shows the estimates of the basic parameters (treatment effects relative to the reference treatment, placebo) obtained from univariate network meta-analyses, bivariate analyses for all three combinations of pairs of outcomes and the trivariate analysis. The results are similar across all analyses, and conclusions from univariate and multivariate analyses are the same. This is disappointing because multivariate analyses have not resulted in more precise inference. The entries of ${\bf \hat{\Sigma}}_\beta$ and ${\bf \hat{\Sigma}}_\omega$  are shown in Table \ref{tab-ms-results2}. The positive estimates obtained for the unknown variance components suggest that this example exhibits some between-study heterogeneity and inconsistency. In order to assess the impact of the unknown variance components, we also fitted the consistent model  and the common-effect and consistent model (results not shown) using all three outcomes ($p=3$). On average, the standard errors of the fifteen basic parameters from the full  model are 35\% greater (range: 13\% to 84\%) than those from the consistent model, which in turn   are 58\% (range: 8\% to 128\%) greater than those from the common-effect and consistent model. Both the between-study heterogeneity and inconsistency have notable impact.

The multivariate analysis adds to the univariate analyses in two main ways. Firstly, the finding that the multivariate analysis is in good agreement with the univariate analyses is a particularly important finding for treatment effects on MRI where a substantial proportion of data were missing. It has been demonstrated by Kirkham {\it et al}. (2012) that a multivariate approach to meta-analysis can help obtain more accurate estimates in the presence of outcome reporting bias. Hence the multivariate analysis reduces concerns that this univariate analysis is affected by reporting bias. Secondly joint inferences for all three outcomes are possible under the multivariate model. For example, and as we might anticipate, in our example the estimated log annualised relapse rate ratios and log disability progression odds ratios are highly positively correlated; from $\mbox{Var}(\hat{{\bm \delta}})$ in our three dimensional multivariate meta-analysis, the correlations between the five pairs of estimated basic parameters for these two outcomes are all between 0.63 and 0.75. Medical decision making based jointly on these two outcomes should take this high positive correlation into account, and this is only possible by using a multivariate approach. For example, a formal decision analysis involving these two outcomes should be based on their joint distribution rather than their two marginal distributions.

\section{Discussion}
We have proposed a new model for dealing with both multiple treatment contrasts and multiple outcomes, to provide a  framework for conducting multivariate network meta-analysis. By using a matrix-based method of moments estimator, our methodology naturally builds on previous work (such as the well-known DerSimonian and Laird approach) and is computationally very fast, relative to other potential estimation approaches such as REML or MCMC; this is especially the case in very high dimensions and so our methodology is particularly advantageous for ambitious analyses of this type. The main disadvantage is that, as a necessary consequence of its semi-parametric nature, the method of moments is not based on sufficient statistics and so is not fully efficient. The loss in efficiency relative to maximum likelihood estimation awaits  investigation but we anticipate that this will be less serious for inferences about the average effects than the unknown variance components. Furthermore the within-study normal approximations used in our model are not necessarily very accurate even in moderately sized studies.

Since our analysis uses a general design matrix, the modelling may easily be extended by adding study level covariates to describe and  fit multivariate network meta-regressions. In the network meta-analysis setting these regressions have the potential to explain the reasons for inconsistency and model multiple dose level responses. Our method of moments estimation can be combined with  approaches that `inflate' confidence intervals from a frequentist random effects meta-analysis (Hartung and Knapp, 2001; Jackson and Riley, 2014).

In conclusion, we have developed a new model and estimation method for multivariate network meta-analysis, which can describe multiple treatments and multiple correlated outcomes.


\section*{Acknowledgements}
 DJ, IRW and ML are (or were) employed by the UK Medical Research Council [Unit Programme number U105260558].
SB was supported by the Medical Research Council (MRC) Methodology Research Programme [New Investigator Research Grant MR/L009854/1].


\newpage

\section*{References}

Achana, F. A., Cooper, N. J., Bujkiewicz, S., Hubbard, S. J., Kendrick, D., Jones, D. R. and Sutton, A. J. (2014).
Network meta-analysis of multiple outcome measures accounting for borrowing of information across outcomes.
{\it BMC Medical Research Methodology} {\bf 14,} 92.\\

\noindent Bujkiewicz, S., Thompson, J. R., Riley, R. D. and Abrams, K. R. (2016).
Bayesian meta-analytical methods to incorporate multiple surrogate endpoints in drug development process.
{\it Statistics in Medicine} {\bf 35,} 1063-–1089.\\

\noindent Chen, H., Manning, A. K. and Dupuis, J. (2012). A method of moments estimator for random effect multivariate meta-analysis. {\it Biometrics} {\bf 68,} 1278-–1284.
\\

\noindent Dersimonian, R. and   Laird, N. (1986). Meta-analysis in clinical trials. {\it Controlled Clinical Trials} {\bf 7,} 177--188.
\\

\noindent Efthimiou, O., Mavridis, D., Cipriani, A., Leucht, S., Bagos, P. and Salanti, G. (2014).
An approach for modelling multiple correlated outcomes in a network of interventions using
odds ratios.
{\it Statistics in Medicine} {\bf 33,} 2275-–2287.
\\

\noindent Efthimiou O, Mavridis D, Riley RD, Cipriani A and Salanti G. (2015)
Joint synthesis of multiple correlated outcomes in networks of interventions.
{\it Biostatistics} {\bf 16,} 84--97.
\\

\noindent Hartung, J. and Knapp, G. (2001). On tests of the overall treatment effect in meta-analysis with normally distributed responses. {\it Statistics in Medicine} {\bf 20,} 1771--1782.
\\

\noindent Henderson, H. V. and Searle, S. R. (1981). The vec-permutation matrix, the vec operator and Kronecker products: a review. {\it Linear and Multilinear Algebra} {\bf 9,} 271–-288.
\\

\noindent Higgins, J.P.T., Jackson, D., Barrett, J.K., Lu, G., Ades, A.E. and White, I.R. (2012). Consistency and inconsistency in network meta-analysis: concepts and
models for multi-arm studies. {\it Research Synthesis Methods} {\bf 3,} 98--110.
\\

\noindent Hong, H., Fu, H., Price, K. L. and Carlin, B. P. (2015)
Incorporation of individual-patient data in network meta-analysis for multiple continuous endpoints, with application to diabetes treatment.
{\it Statistics in Medicine} {\bf 34,} 2794--2819.
\\

\noindent Hong, H., Chu H., Zhang J. and Carlin B.P.  (2016)
A Bayesian missing data framework for generalized multiple outcome mixed treatment comparisons.
{\it Research Synthesis Methods} {\bf 7,} 6--22.
\\

\noindent Jackson, D., Riley, R. and White, I. R. (2011). Multivariate meta-analysis: potential and promise (with discussion). {\it Statistics in Medicine} {\bf 30,} 2481–-2510.
\\

\noindent Jackson, D., White, I.R. and Riley, R.D. (2013). A matrix-based method of moments for fitting the multivariate random effects model for meta-analysis and meta-regression. {\it Biometrical Journal} {\bf 55,} 231--245.
\\

\noindent Jackson, D. and Riley, R. (2014).  A refined method for multivariate meta-analysis and meta-regression. {\it Statistics in Medicine} {\bf 33,} 541--554.
\\

\noindent Jackson, D., Law, M., Barrett, J.K., Turner, R., Higgins, J.P.T., Salanti, G. and White, I.R. (2016). Extending DerSimonian and Laird's methodology to perform network meta-analyses with random inconsistency effects. {\it Statistics in Medicine} {\bf 35}, 819--839.
\\

\noindent Kirkham, J. J., Riley, R. D. and Williamson, P. R. (2012).
A multivariate meta-analysis approach for reducing the impact of outcome reporting bias in systematic reviews.
{\it Statistics in Medicine} {\bf 31}, 2179-–2195.
\\

\noindent Kulinskaya, E., Dollinger, M.B. and Bj{\o}rkest{\o}l, K. (2011). Testing for Homogeneity in Meta-Analysis I. The One-Parameter Case: Standardized Mean Difference. {\it Biometrics} {\bf 67,} 203--212.
\\

\noindent Law, M., Jackson, D., Turner, R., Rhodes, K. and Viechtbauer, W. (2016).  Two new methods to fit models for network meta-analysis with random inconsistency effects
{\it BMC Medical Research Methodology} {\bf 16}, 87
\\

\noindent Lu, G. and Ades, A. (2004). Combination of direct and indirect evidence in mixed treatment comparisons. {\it Statistics in Medicine} {\bf 23,} 3105--3124.
\\

\noindent Lu, G. and Ades, A. (2006). Assessing evidence consistency in mixed treatment comparisons. {\it Journal of the American Statistical Association} {\bf 101,} 447--459.
\\

\noindent Nikolakopoulou, A., Chaimani, A., Veroniki, A., Vasiliadis, H.S., Schmid, C.H. and Salanti, G. (2014). Characteristics of Networks of Interventions: A Description of a Database of 186 Published Networks. {\it Plos One} {\bf 9}, 1: e86754.
\\

\noindent Piepho, H.P., Williams, E.R. and Madden, L.V. (2012). The Use of Two-Way Linear Mixed Models in Multitreatment
Meta-Analysis. {\it Biometrics} {\bf 68,} 1269--1277.
\\

\noindent Riley, R.D.,  Thompson, J. R. and Abrams, K. R. (2008).
An alternative model for bivariate random-effects meta-analysis
when the within-study correlations are unknown.
{\it Biostatistics} {\bf 9}, 172-–186.
\\

\noindent  Riley, R.D., Price, M.J., Jackson, D., Wardle, M., Gueyffier, F., Wang, J., Staessen, J.A. and White, I.R. (2015). Multivariate meta-analysis using individual participant data. {\it Research Synthesis Methods} {\bf 6,} 157–-174.
\\

\noindent Sormani, M.P., Bonzano, L., Roccatagliata, L.,  Mancardi, G.L., Uccelli, A. and Bruzzi, P. (2010). Surrogate endpoints for EDSS worsening
in multiple sclerosis. A meta-analytic approach. {\it Neurology} {\bf 75,} 302--309.
\\

\noindent Seaman, S., Galati, J., Jackson, D. and  Carlin, J. (2013). What Is Meant by ``Missing at Random?''. {\it Statistical Science } {\bf 28}, 257-268.
\\

\noindent Searle, S.R. (1971). Linear Models. Wiley. New York.
\\

\noindent Veroniki, A., Vasiliadis, H.S., Higgins, J.P. and Salanti G. (2013). Evaluation of inconsistency in networks of interventions. {\it International Journal of Clinical Epidemiology} {\bf 42,} 332--345.
\\

\noindent Wei, Y. and Higgins, J.P.T. (2013). Estimating within-study covariances in multivariate meta-analysis with multiple outcomes. {\it Statistics in Medicine} {\bf 32,} 1191–-1205.

\newpage

\section*{Supplementary Materials}
\section*{Multivariate estimation}
 \subsection*{An important result}
 In order to evaluate the expectations required, we will need to be able to compute expressions of the form $\mbox{btr}({\bf A}({\bf M} \otimes {\bf \Sigma}){\bf B})$, where ${\bf A}$ and ${\bf B}$ are $np \times np$ matrices, ${\bf M}$ is an $n \times n$ matrix and ${\bf \Sigma}$ is a $p \times p$ matrix. We continue to use the notation ${\bf A}_{i,j}$ to denote the $i$th by $j$th block of ${\bf A}$, where  these blocks are $p \times p$ matrices. For any three $np \times np$ matrices ${\bf A}$,  ${\bf B}$ and ${\bf C}$, we have
\[
({\bf ACB})_{k,l} = \sum\limits_{i=1}^{n}\sum\limits_{j=1}^{n} {\bf A}_{k,i } {\bf C}_{i,j } {\bf B}_{j,l }
\]
This is just the law of matrix multiplication applied to blocks. Then taking ${\bf C} = {\bf M} \otimes {\bf \Sigma} $ so that from the definition of the Kronecker product, ${\bf C}_{i,j} = m_{ij} {\bf \Sigma}$, we have
\[
({\bf A({\bf M} \otimes {\bf \Sigma})B})_{k,l} =  \sum\limits_{i=1}^{n}\sum\limits_{j=1}^{n} m_{ij} {\bf A}_{k,i } {\bf \Sigma} {\bf B}_{j,l }
\]
To obtain the block trace, we sum the matrices along the main diagonal. Hence to obtain the block trace we take $l=k$ to obtain the matrices along the main diagonal and sum over $k$ so obtain
\begin{equation}
\label{ap1}
\mbox{btr}({\bf A}({\bf M} \otimes {\bf \Sigma}){\bf B}) = \sum\limits_{i=1}^{n}\sum\limits_{j=1}^{n} \sum\limits_{k=1}^{n} m_{ij} {\bf A}_{k,i } {\bf \Sigma} {\bf B}_{j,k }
\end{equation}
The use of equation (\ref{ap1}), with the appropriate matrices, almost immediately results in the expected values required in section 4.

  \subsection*{The estimating equations }
  In this section we prove the results given in section 4 of the main paper. We do not redefine all quantities or give the size of all matrices and vectors, see the main paper for these details.  As in  the univariate approach of Jackson {\it et al.} (2016), we will base our estimation on the two quantities
 $\mbox{btr} ({\mathbf Q}) $
  and $  \sum\limits_{d=1}^D\mbox{btr} ({\mathbf Q}_d)$
  where $D$ is the number of different designs. We match these quantities to their expectations to estimate the unknown variance parameters. We therefore need to evaluate  $\mbox{E}[\mbox{btr} ({\mathbf Q})] $ and $\mbox{E}[\mbox{btr}({\mathbf Q}_d)]$.

 \subsubsection*{Evaluating $\mbox{E}[\mbox{btr} ({\mathbf Q})] $ and deriving the first estimating equation}
As in Jackson {\it et al.} (2013), by direct calculation we have that ${\mathbf W} {\mathbf H} {\mathbf W}^{-1} = {\mathbf H}^T$ and $(({\mathbf I}_{np} - {\mathbf H})^T)^2= ({\mathbf I}_{np} - {\mathbf H})^T$; if ${\mathbf W}$ is not invertible because outcome data are missing then we can justify the use of the identity ${\mathbf W} {\mathbf H} {\mathbf W}^{-1} = {\mathbf H}^T$ and the expectation that follows in the limit, where the precision $p$ attributed to missing data tends towards zero from above, $p \rightarrow 0^+$ (Jackson {\it et al.}, 2013). Furthermore we can use the identity $ {\mathbf W} = {\mathbf S}^{-1}$ in this limit. We also have that ${\mathbf Y} - \hat{{\mathbf Y}} = ({\mathbf I}_{np} - {\mathbf H}){{\mathbf Y}} $ and  $\mbox{E}[{\mathbf Y} - \hat{{\mathbf Y}}]= {\mathbf 0}$. Hence from the definition of ${\mathbf Q}$  we have $\mbox{E}[{\mathbf Q}] = {\mathbf W}\mbox{Var}[{\mathbf Y}-\hat{{\mathbf Y}}] {\mathbf R}$.  From these results,  taking the variance of ${\mathbf Y}$ from model (3) of the main paper,  we can evaluate
\[
\mbox{E}[{\mathbf Q}]={\mathbf A}({{{\bf M_1} \otimes {\bf \Sigma}_{\beta} +  {{\bf M_2}}\otimes {\bf \Sigma}_{\omega} }}) {\mathbf B} + {\mathbf B}
\]
where
\[
{\mathbf A} = ({\mathbf I}_{np}-{\mathbf H})^T {\mathbf W}
\]
and
\[
{\mathbf B} = ({\mathbf I}_{np} - {\mathbf H})^{T} {\mathbf R}
\]
Here ${\mathbf A}$ and ${\mathbf B}$ are known $np \times np$ matrices.  For estimation purposes we require  $  \mbox{E}[\mbox{btr}({\mathbf Q})] =\mbox{btr} (\mbox{E}[{\mathbf Q}])$. We  write ${\mathbf A}_{i,j}$ and ${\mathbf B}_{i,j}$ to mean the $i$th by $j$th blocks of  ${\mathbf A}$ and ${\mathbf B}$ respectively, so that ${\mathbf A}_{i,j}$ and ${\mathbf B}_{i,j}$ are both $p \times p$ matrices. Then, using (\ref{ap1}),  we have
\[
\mbox{E}[\mbox{btr}({\mathbf Q})]=
\sum\limits_{i=1}^{n}\sum\limits_{j=1}^{n}\sum\limits_{k=1}^{n} m_{1i j} {\mathbf A}_{k,i} {\bf \Sigma}_\beta {\mathbf B}_{j,k}+
\]
\[
\sum\limits_{i=1}^{n}\sum\limits_{j=1}^{n}\sum\limits_{k=1}^{n} m_{2i j} {\mathbf A}_{k,i} {\bf \Sigma}_\omega {\mathbf B}_{j,k} + \mbox{btr}({\mathbf B})
\]

\subsubsection*{Evaluating $\mbox{E}[\mbox{btr} ({\mathbf Q}_d)] $ and deriving the second estimating equation}
Then we follow very similar, but much simpler, arguments as in the previous section to derive the result that we require. We define design specific hat matrices
\begin{equation}
\label{ap2}
{\mathbf H}_d = {\mathbf X}_d ({\mathbf X}_d^T {\mathbf W}_d {\mathbf X}_d)^{-1} {\mathbf X}_d^{T} {\mathbf W}_d
\end{equation}
 and also design specific $p n_d \times p n_d$ ${\mathbf A}$ and ${\mathbf B}$ matrices
\[
{\mathbf A}_d = ({\mathbf I}_{p n_d}-{\mathbf H}_d)^T {\mathbf W}_d
\]
and
\[
{\mathbf B}_d = ({\mathbf I}_{p n_d} - {\mathbf H}_d)^{T} {\mathbf R}_d
\]
In equation (\ref{ap2}) we take the matrix inverse to be the Moore-Penrose pseudoinverse. This is because, in the presence of missing outcome data, the design-specific regression corresponding to this hat matrix may not be identifiable (for example, if studies of a particular design do not provide data for one or more of the  outcomes). In such instances this design may still provide information about some of the unknown between-study variance components and so it is not desirable to exclude the design from this part of the estimation procedure. By computing (\ref{ap2}) using this pseudoinverse we obtain a suitable hat matrix (Searle, 1971; page 221, his equations 126 and 127). Furthermore all the necessary properties of the hat matrix are retained when using the pseudoinverse when computing (\ref{ap2}) and we retain unbiased fitted values (Searle, 1971; page 181).

 Following a simpler version of the arguments in the previous section and the main paper, taking the variance of ${\mathbf Y}_d$ from model (5) of the main paper, and upon applying the vec operator, we obtain
\begin{equation}
\label{ap3}
\mbox{vec}(\mbox{E}[\mbox{btr}({\mathbf Q}_d)]) = {\mathbf C}_d \mbox{vec}({\bf \Sigma}_\beta)  + {\mathbf E}_d
\end{equation}
where
\[
{\mathbf C}_d = \sum\limits_{i=1}^{n_d}\sum\limits_{j=1}^{n_d}\sum\limits_{k=1}^{n_d} m^d_{1 ij} {\mathbf B}_{d, j,k}^T \otimes {\mathbf A}_{d, k,i}
\]
and
\[
{\mathbf E}_d = \mbox{vec}(\mbox{btr}({\bf B}_d))
\]
We then sum equation (\ref{ap3}) across all designs in order to obtain
\[
\mbox{vec}\left(\mbox{E}[\sum\limits_{d=1}^D \mbox{btr}({\mathbf Q}_d)]\right) = \left(\sum\limits_{d=1}^D {\mathbf C}_d\right) \mbox{vec}({\bf \Sigma}_\beta)  + \sum\limits_{d=1}^D {\mathbf E}_d
\]

\subsection*{Special cases of the estimation procedure (an extended version of section 4.5)}
The proposed method reduces to two previous methods in special cases. If all studies are two arm studies (and so provide a single contrast) and consistency is assumed then the proposed method reduces to the matrix based method for multivariate meta-regression  (Jackson {\it et al.}, 2013). This is because we then have ${\bf \Sigma}_\omega={\bf 0}$, so that the second triple sum in our expression for $\mbox{E}[\mbox{btr}({\mathbf Q})]$ is zero; furthermore the first triple summation in this expression can be reduced to a double summation, because ${\bf M_1}$ is an identity matrix for multivariate meta-regression (Jackson {\it et al.}, 2013; their equation A.1.).

Furthermore the proposed multivariate method also reduces to the univariate DerSimonian and Laird method for network meta-analysis (Jackson {\it et al.}, 2016) when $p=1$. This is because, in one dimension, the ${\mathbf Q}$ matrices all reduce to the $Q$ random scalars used in the estimation procedure suggested by  Jackson {\it et al.} (2016). This can be shown by replacing the block trace operator with the more familiar trace of a matrix (btr is the trace when $p=1$) in the definition of the ${\mathbf Q}$ matrices and using the identity $\mbox{tr}({\bf AB}) =  \mbox{tr}({\bf BA})$. These two special cases are in turn generalisations of methods such as that proposed by DerSimonian and Laird (1986).

 There is however one caveat when stating that the new multivariate method reduces to the univariate method proposed by  (Jackson {\it et al.}, 2016) when $p=1$. This is because the account of Jackson {\it et al.} (2016) does not mention the possibility of missing outcome data and so we have implicitly taken all data to be observed in the argument used in the previous paragraph.

\section*{Example of matrices $\mathbf{M_1}$ and $\mathbf{M_2}$}
We provide a concrete example of matrices $\mathbf{M_1}$ and $\mathbf{M_2}$, in order to clarify how they are computed. We take such an example from Law {\it et al.} (2016) which comprises thirteen studies with the following study designs: AB, BC, BC, BC, BC, BC, BD, BD, CD, CD, ABD, BCD, BCD. This is the same type of network as used in the simulation study below. The two matrices for this example are given explicitly below, where we can see that these matrices contain blocks that are comprised of blocks of  $\mathbf{P}_{c_{d}}$, where in $\mathbf{M_1}$ the blocks are formed by studies and in $\mathbf{M_2}$ the blocks are formed by designs.

\vspace{15mm}

\begin{center}

\setcounter{MaxMatrixCols}{20}
$\mathbf{M_1} = \left( \begin{array}{c|c|c|c|c|c|c|c|c|c|cc|cc|cc}
       			   1 & 0 & 0 & 0 & 0 & 0 & 0 & 0 & 0 & 0 & 0 & 0 & 0 & 0 & 0 & 0 \\
       			   \hline
                   0 & 1 & 0 & 0 & 0 & 0 & 0 & 0 & 0 & 0 & 0 & 0 & 0 & 0 & 0 & 0 \\
                   \hline
                   0 & 0 & 1 & 0 & 0 & 0 & 0 & 0 & 0 & 0 & 0 & 0 & 0 & 0 & 0 & 0 \\
                  \hline
                   0 & 0 & 0 & 1 & 0 & 0 & 0 & 0 & 0 & 0 & 0 & 0 & 0 & 0 & 0 & 0 \\
				  \hline
                   0 & 0 & 0 & 0 & 1 & 0 & 0 & 0 & 0 & 0 & 0 & 0 & 0 & 0 & 0 & 0 \\
                   \hline
                   0 & 0 & 0 & 0 & 0 & 1 & 0 & 0 & 0 & 0 & 0 & 0 & 0 & 0 & 0 & 0 \\
                   \hline
                   0 & 0 & 0 & 0 & 0 & 0 & 1 & 0 & 0 & 0 & 0 & 0 & 0 & 0 & 0 & 0 \\
                   \hline
                   0 & 0 & 0 & 0 & 0 & 0 & 0 & 1 & 0 & 0 & 0 & 0 & 0 & 0 & 0 & 0 \\
                   \hline
                   0 & 0 & 0 & 0 & 0 & 0 & 0 & 0 & 1 & 0 & 0 & 0 & 0 & 0 & 0 & 0 \\
                   \hline
                   0 & 0 & 0 & 0 & 0 & 0 & 0 & 0 & 0 & 1 & 0 & 0 & 0 & 0 & 0 & 0 \\
                   \hline
                   0 & 0 & 0 & 0 & 0 & 0 & 0 & 0 & 0 & 0 & 1 & \frac{1}{2} & 0 & 0 & 0 & 0 \\
                   0 & 0 & 0 & 0 & 0 & 0 & 0 & 0 & 0 & 0 & \frac{1}{2} & 1 & 0 & 0 & 0 & 0 \\
                   \hline
                   0 & 0 & 0 & 0 & 0 & 0 & 0 & 0 & 0 & 0 & 0 & 0 & 1 & \frac{1}{2} & 0 & 0 \\
                   0 & 0 & 0 & 0 & 0 & 0 & 0 & 0 & 0 & 0 & 0 & 0 & \frac{1}{2} & 1 & 0 & 0 \\
                   \hline
                   0 & 0 & 0 & 0 & 0 & 0 & 0 & 0 & 0 & 0 & 0 & 0 & 0 & 0 & 1 & \frac{1}{2} \\
                   0 & 0 & 0 & 0 & 0 & 0 & 0 & 0 & 0 & 0 & 0 & 0 & 0 & 0 & \frac{1}{2} & 1 \\
    \end{array}
    \right)$

\vspace{15mm}

\setcounter{MaxMatrixCols}{20}
$\mathbf{M_2} = \left( \begin{array}{c|ccccc|cc|cc|cc|cccc}
                   1 & 0 & 0 & 0 & 0 & 0 & 0 & 0 & 0 & 0 & 0 & 0 & 0 & 0 & 0 & 0 \\
                   \hline
                   0 & 1 & 1 & 1 & 1 & 1 & 0 & 0 & 0 & 0 & 0 & 0 & 0 & 0 & 0 & 0 \\
                   0 & 1 & 1 & 1 & 1 & 1 & 0 & 0 & 0 & 0 & 0 & 0 & 0 & 0 & 0 & 0 \\
                   0 & 1 & 1 & 1 & 1 & 1 & 0 & 0 & 0 & 0 & 0 & 0 & 0 & 0 & 0 & 0 \\
                   0 & 1 & 1 & 1 & 1 & 1 & 0 & 0 & 0 & 0 & 0 & 0 & 0 & 0 & 0 & 0 \\
                   0 & 1 & 1 & 1 & 1 & 1 & 0 & 0 & 0 & 0 & 0 & 0 & 0 & 0 & 0 & 0 \\
                   \hline
                   0 & 0 & 0 & 0 & 0 & 0 & 1 & 1 & 0 & 0 & 0 & 0 & 0 & 0 & 0 & 0 \\
                   0 & 0 & 0 & 0 & 0 & 0 & 1 & 1 & 0 & 0 & 0 & 0 & 0 & 0 & 0 & 0 \\
                   \hline
                   0 & 0 & 0 & 0 & 0 & 0 & 0 & 0 & 1 & 1 & 0 & 0 & 0 & 0 & 0 & 0 \\
                   0 & 0 & 0 & 0 & 0 & 0 & 0 & 0 & 1 & 1 & 0 & 0 & 0 & 0 & 0 & 0 \\
                   \hline
                   0 & 0 & 0 & 0 & 0 & 0 & 0 & 0 & 0 & 0 & 1 & \frac{1}{2} & 0 & 0 & 0 & 0 \\
                   0 & 0 & 0 & 0 & 0 & 0 & 0 & 0 & 0 & 0 & \frac{1}{2} & 1 & 0 & 0 & 0 & 0 \\
                   \hline
                   0 & 0 & 0 & 0 & 0 & 0 & 0 & 0 & 0 & 0 & 0 & 0 & 1 & \frac{1}{2} & 1 & \frac{1}{2} \\
                   0 & 0 & 0 & 0 & 0 & 0 & 0 & 0 & 0 & 0 & 0 & 0 & \frac{1}{2} & 1 & \frac{1}{2} & 1 \\
                   0 & 0 & 0 & 0 & 0 & 0 & 0 & 0 & 0 & 0 & 0 & 0 & 1 & \frac{1}{2} & 1 & \frac{1}{2} \\
                   0 & 0 & 0 & 0 & 0 & 0 & 0 & 0 & 0 & 0 & 0 & 0 & \frac{1}{2} & 1 & \frac{1}{2} & 1 \\
\end{array}
\right)$
\end{center}

\end{document}